\def\rfr#1{Eq. (\ref{#1})}

\def\dert#1#2{\frac{{{d}}{#1}}{{{d}}{#2}}}              % derivate parziali e totali prima e seconda

\def\virg#1{``#1''}

\def\eqi{\begin{equation}}
\def\eqf{\end{equation}}
\def\eqia{\begin{eqnarray}}
\def\eqfa{\end{eqnarray}}
\def\rp#1#2{{#1\over#2}} \def\lb#1{\label{#1}}

\def\kp{\hat{k}_p}
\def\kq{\hat{k}_q}
\def\kh{\hat{k}_h}
\def\Om{\mathit{\Omega}}
\def\Ps{\mathit{\Psi}}
\def\bds#1{\boldsymbol{#1}}
\def\kap{\bds{\hat{k}}}
\def\kx{\hat{k}_x}
\def\ky{\hat{k}_y}
\def\kz{\hat{k}_z}
\def\si{\sin i}
\def\ci{\cos i}

\def\sO{\sin\Om}
\def\cO{\cos\Om}

\documentclass[%
 reprint,
showpacs,showkeys,
 amsmath,amssymb,
 aps,
]{revtex4-1}
\usepackage{graphicx}% Include figure files
\usepackage{dcolumn}% Align table columns on decimal point
\usepackage{bm}% bold math
\usepackage{url}
\usepackage{amsmath,amsthm,amscd,amssymb}
\usepackage{latexsym,wasysym}
\usepackage{graphicx,epsfig}

\RequirePackage{color}

\begin{document}

\title{Perturbed stellar motions  around the rotating black hole in Sgr A$^{\ast}$ for a generic orientation of its spin axis}

\author{L. Iorio}
\altaffiliation{Ministero dell'Istruzione, dell'Universit\`{a} e della Ricerca (M.I.U.R.)-Istruzione. Fellow of the Royal Astronomical Society (F.R.A.S.).
 International Institute for Theoretical Physics and
Advanced Mathematics Einstein-Galilei. Permanent address: Viale Unit$\grave{\rm a}$ di Italia 68
70125 Bari (BA), Italy.}
\email{lorenzo.iorio@libero.it}

\begin{abstract}
Empirically determining the averaged variations of the orbital parameters of the stars orbiting the Supermassive Black Hole (SBH) hosted by the Galactic center (GC) in Sgr A$^{\ast}$ is, in principle, a valuable tool to  test the General Theory of Relativity (GTR), in  regimes far stronger than those tested so far, and certain key predictions of it like the no-hair theorems. We analytically work out the long-term variations of all the six osculating Keplerian orbital elements of a test particle orbiting a nonspherical, rotating body with quadrupole moment $Q_2$ and angular momentum $\bds S$ for a generic spatial orientation of its spin axis $\kap$.  This choice is motivated by the fact that, basically, we  do not know the position in the sky of the spin axis of the SBH in Sgr A$^{\ast}$ with sufficient accuracy. We apply our results to  S2, which is the closest star  discovered so far having an orbital period $P_{\rm b}=15.98$ yr, and to a hypothetical closer star $X$ with $P_{\rm b}=0.5$ yr. Our calculations are quite general, not being related to any specific parameterization of $\kap$, and can be applied also to astrophysical binary systems, stellar planetary systems, and planetary satellite geodesy in which different reference frames, generally not aligned with the primary's rotational axis, are routinely used.
\end{abstract}

%\keywords{Classical general relativity;	Physics of black holes; Experimental tests of gravitational theories;  Harmonics of the gravity potential field; %Satellite orbits; Black holes}

\pacs{04.20.-q, 04.70.-s, 04.80.Cc, 91.10.Qm, 91.10.Sp, 97.60.Lf}

\maketitle

\section{Introduction}
 There is nowadays wide consensus \cite{Genzel,nigri,Ghez} that the Galactic Center (GC)  hosts a Supermassive Black Hole (SBH) \cite{Wol,Falke} whose  position coincides with that of the radio-source Sagittarius A$^{\ast}$ (Sgr A$^{\ast}$) \cite{Bal,Reid07} at $d=8.28\pm 0.44$ kpc  from us  \cite{Gille}; for a popular overview of such an object, see, e.g., Ref. \cite{Melia07}. The Galactic SBH is surrounded by a number of recently detected main-sequence stars of spectral class B \cite{Pau,Gille}. They have been revealed and tracked in the
near infrared since 1992 at the 8.2 m Very Large Telescope (VLT) on Cerro Paranal, Chile and the 3.58 m New Technology Telescope (NTT)
on La Silla, Chile \cite{Ecka}, and since 1995 at the Keck 10 m telescope on Mauna Kea, Hawaii \cite{Ghez98}. They are dubbed SN, or S0-N in the Keck nomenclature, where N is a progressive order number. Their relatively fast orbital motions, characterized by orbital periods $P_{\rm b}\gtrsim 16$ yr, high eccentricities $e\gtrsim  0.2$, random orientations $i$ of their orbits in the sky and average distances from the SBH $\overline{r} \gtrsim 2\times 10^4 r_g$, where $r_g$ denotes the Schwarzschild radius, allowed to dynamically infer a mass of about $M\approx 4\times 10^6 M_{\odot}$ \cite{Ghez,Gille,Gille2} for it.

The direct access to  such S/S0 stars, and of other closer objects which may hopefully be discovered in the future, has induced several researchers to investigate various predictions that the General Theory of Relativity (GTR) directly makes for their orbital motions along with other competing effects from standard Newtonian gravity which may mask the relativistic ones \cite{Jaro,Fragile,Rubi,Weinb,Kran,Nuci,Will08,Preto,Khan,Merr010,Iorio011,Sade011}. Concerning several effects related to propagating electromagnetic waves in connection with the stellar orbital motions like, e.g., relativistic redshifts, see Ref. \cite{Zuck06,Ange010a,Ange010b,Ange011}. In fact, although the currently known stars, in a strict sense, cannot probe the  strong field regime of GTR because of their relatively large distance from the SBH, on the other hand they yield a unique opportunity to put on the test GTR in the strongest field regime ever probed so far. Indeed, even in the double binary pulsar PSR J0737-3039A/B \cite{pulsar,Lyne} $r_g/\overline{r}$ is one order of magnitude smaller than for S2, which is the closest SBH star discovered so far \cite{Gille}.

In this paper we
 analytically work out the averaged variations of all the six standard osculating Keplerian parameters of a test particle caused by the rotation of the central object endowed with angular momentum $\bds S=S\kap$ and quadrupole moment $Q_2$. Note that the stars orbiting the SBH can  safely be considered test particles: their masses are about $m\lesssim 10^{-5} M$, and relativistic corrections to their internal structures are assumed to be too small to yield noticeable effects on their orbital motions. No assumptions about any specific spatial orientation for  $\kap$ are made. Thus, our calculations are not restricted to a particular reference frame, and are valid also for different scenarios like, e.g., stellar planetary systems and planetary satellite geodesy in which natural and/or artificial test bodies are employed for testing GTR. Moreover, in order to keep our calculations as general as possible, we will not adopt any particular representation for $\kap$ in terms of specific angular variables in the sense that we will refer it to the global reference frame adopted; for a different approach, see Ref. \cite{Will08} in which $\kap$ is referred to the orbital plane of each star. Concerning the SBH in Sgr A$^{\ast}$, the orientation of its spin axis is substantially unknown, despite the attempts by different groups  \cite{spin1,spin2,mmvlbi2} to constrain it using different parameterizations which yielded quite loose bounds. A strategy to partially overcome such an obstacle have been recently put forward in Ref. \cite{Hio}; it is based on the possible observation of the apparent shape of the shadow cast by the BH on the plane of the sky, and would allow the measurement of $S$ and the angle $i^{'}$ between $\kap$ and the line-of-sight.

The GTR  prediction for the standard 1PN periastron precession,  which is analogous to Mercury's well known perihelion precession of $42.98$ arscec cty$^{-1}$ and is independent of $\kap$, amounts to about \eqi\dot\omega_{\rm S2}^{(\rm 1PN)}=45\pm 10\ {\rm arcsec\ yr}^{-1}\lb{yuo}\eqf for S2; the quoted uncertainty comes from the errors in the parameters of both the SBH and S2 entering the GTR formula: they are displayed in Table \ref{tavola}. The result of \rfr{yuo}, computed in a frame with the SBH at its origin, corresponds to a precession \textcolor{black}{$\dot\xi$ as seen from the Earth of} $\dot\xi_{\rm S2} = 27\pm 6$ microarcseconds per year ($\mu$as yr$^{-1}$ in the following). At present, it is still undetectable from the currently available direct astrometric measurements in terms of right ascension $\alpha$ and declination $\delta$ which barely cover just one full orbital period of S2.
Indeed, according to Table 1 of Ref. \cite{Gille}, the present-day error in  the periastron  is $\sigma_{\omega}=0.81\ {\rm deg}=2916$ arcsec over about 16 yr, from which an uncertainty of about $\sigma_{\dot\omega} \simeq 182$ arcsec yr$^{-1}$ in the periastron precession may naively be inferred: it corresponds to a limiting accuracy  of $\sigma_{\dot\xi}=110$ $\mu$as yr$^{-1}$  in monitoring angular rates as seen from the Earth. As we will see, the sizes of the other precessions of S2 due to $S$ and $Q_2$ may be smaller by about $2$ and $4-5$ orders of magnitude, respectively for a moderate rotation of the SBH.
Concerning future perspectives, according to Ref. \cite{Eise09} future astrometric measurements of S2 may bring its 1PN periastron rate into the measurability domain; indeed, the periastron advance would  indirectly be inferred from the corresponding apparent position shift in the recorded orbit. Moreover, the ASTrometric and phase-Referenced Astronomy (ASTRA) project \cite{Eisner}, to be applied to the Keck interferometer, should be able to monitor stellar orbits with an accuracy of \cite{Pott} $\sigma_{\Delta\xi}\approx 30$ $\mu$as as seen from the Earth.   The  GRAVITY instrument \cite{Gille010}, devoted to enhance the capabilities of the VLT interferometer (VLTI),  aims to reach an accuracy of $\sigma_{\Delta\xi}\approx 10$ $\mu$as \cite{Gille010} in measuring astrometric shifts $\Delta\xi$ as seen from the Earth, which, among other things, would allow exploration of the innermost stable circular orbits around the SBH  \cite{Vinc011}.

About testing GTR in the SBH scenario, we make the following general considerations. In order to meaningfully compare theoretical predictions for a given effect to its empirically determined counterpart, we need to know some of the key ambient parameters entering the predictions independently from the effects themselves we are looking for. In the specific case, the mass $M$, the spin $\bds S$ and the quadrupole $Q_2$ of the SBH should be known, if possible, independently of the precessions  we are going to consider.
Concerning the SBH mass $M$,  the values which we presently have for it can be thought as inferred from the third Kepler law used in conjunction with the directly measured orbital period $P_{\rm b}$, and the semimajor axis $a$ empirically determined by modeling the recorded stellar orbit in the plane of the sky with an ellipse (see Fig. 2 of Ref. \cite{Gille}). Such a determination of $M$ would be, in principle, \virg{imprinted} by GTR itself since it implies a correction to the third Kepler law, but it is far too small with respect to the present-day accuracy in determining $P_{\rm b}$. Indeed, it is $\sigma_{P_{\rm b}}\simeq 10^{-1}$ yr \cite{Ghez,Gille2}, while the 1PN GTR correction to the Keplerian orbital period is \cite{Dam,Soffel} $\Delta P_{\rm b}^{(\rm 1PN)}\propto (3\pi/c^2)\sqrt{GM a}\simeq 10^{-3}$ yr for S2. The same holds also if $M$ is straightforwardly inferred, in a perhaps less transparent manner, as a solve-for quantity from multiparameter global fits of all the stars' data: modeling or not GTR at 1PN level has not yet statistically significant influence in its estimated values, as shown by Table 2 of Ref. \cite{Gille}. We stress that, when such an approach is followed to test GTR, it is intended that different dynamical models, with and without GTR, are fitted to the same data sets to see if statistically significant differences occur in the solve-for estimated parameters.
The quadrupole moment $Q_2$ of the SBH in Sgr A$^{\ast}$ may be measured, e.g., by means of imaging observations  with Very Long
Baseline Interferometry (VLBI) in the strong field regime; see Ref. \cite{bambi,bambib,Joha} for  recent reviews and other proposals.
In regard to the spin $\bds S$ of Sgr A$^{\ast}$, one tries to gain independent information about $S$ from the interpretation  of some measured Quasi-Periodic Oscillations (QPOs)
in the X-ray spectrum of the electromagnetic radiation emitted by the gas orbiting in the accretion disk close to its inner edge \cite{Kato,Genznat}. More recent observations conducted with the Millimeter Very Long Baseline Interferometry (mm-VLBI), probing the immediate vicinity of the horizon, have been able to get information on $S$ \cite{spin2,mmvlbi2}. In interpreting such measurements,  the validity of the Kerr metric \cite{Kerr}  as predicted by GTR is assumed, thus inferring $S$ from, say, the radius of the inner edge extracted from the X-ray diagnostics. It is worth pointing out that the mere fact of obtaining a good fit of the Kerr metric \cite{Kerr}  to a certain empirically determined quantity like, e.g., the X-ray spectrum,  getting a reasonable value for $S$ as a least-square adjustable parameter,  cannot be considered in itself as a test of the rotation-related predictions of GTR, also because other competing mechanisms to explain, say, the QPOs, whose physics is still rather disputed, exist.  Independent empirical determinations of different effects connected with $\bds S$ are required, and the stellar orbital precessions would be just what we need. The greater the number of precessions empirically determined, the greater the number of GTR tests which can be performed. In principle,  more than five precessions are required since $M,S,Q_2$ and two components of $\kap$ must be determined; see also the discussion in Ref. \cite{Will08}. Thus, the need for more than one star is apparent. Such a number of necessary orbital rates may be reduced if some of the aforementioned parameters are somehow reliably obtained from other sources. Of course, also the accuracy with which the precessions can be determined plays a role, in the sense that the previous reasoning holds in the ideal case in which all the three dynamical  effects considered are detectable. Basically, it is the same logic behind the usual tests in the binary systems hosting at least one active radio-pulsar \cite{Damo}. Indeed, in that case the interpretation  of just two empirically determined post-Keplerian effects in terms of their 1PN-GTR predictions is not sufficient since it only allows to obtain the masses $m_1$ and $m_2$ of the system, which are a priori unknown. In the binary pulsar systems the effects which can, actually, be inferred from the data are not limited just to the post-Keplerian periastron precession. Genuine tests of GTR are made when more than two post-Keplerian parameters are empirically determined, and the additional ones are interpreted with GTR by using in their analytical predictions just the previously obtained values for $m_1$ and $m_2$ \cite{Damo}.
%(or from other techniques allowing to unambiguously connect some measured quantities with the radius of the inner edge of the accretion disk. It is by no means %not a trivial task).
%
% Thus, measuring the stellar orbital precessions yields us a tool for performing  genuine tests of GTR. More precisely, we need at least four precessions to %determine $S$, to be compared with the value inferred from X-ray spectroscopy, and $\kap$. It is worth pointing out that the simple fact of having obtained a %good fit of the Kerr metric to the measured X-ray spectrum  getting a reasonable value for $S$  cannot be considered in itself as a test of GTR at all. In %fact, for this purpose we would need to know $S$ independently. Actually, in order to perform a genuine test of GTR, that value of $S$ has to be inserted into %the GTR predictions for some other empirically detectable effects: if the agreement among such theoretical, GTR-based predictions, computed with just that %value of $S$, and the empirical determinations is satisfactorily, then GTR will be considered successfully-and genuinely-tested. Thus, the determination of the %orbital precessions fits just such a scheme, being crucial to put GTR on the test in the BH scenario.
%

The plan of the paper is as follows. In Sec. \ref{due} we review basic facts of standard perturbation theory which will be applied in Sec. \ref{orbisa} to $Q_2$ (Sec. \ref{tre}) and $\bds S$ (Sec. \ref{quattro}). In Sec. \ref{onda} it is briefly remarked that also  gravitational waves with ultralow frequency traveling from the outside could be constrained by the orbital precessions of the stars in Sgr A$^{\star}$. In Sec. \ref{compa} we compare our results to those obtained by Will \cite{Will08}. Numerical evaluations of the effects worked out in Sec. \ref{orbisa} are presented in Sec. \ref{cinque}. Sec. \ref{sei} is devoted to  summarizing our findings and to the conclusions.
\section{Overview of the method adopted}\lb{due}
Here we deal with a generic perturbing acceleration $\bds A$ induced by a given dynamical effect which can be considered as small with respect to the main Newtonian monopole $A_{\rm Newton}=-GM/r^2$, where $G$ is the Newtonian constant of gravitation and $r$ is the mutual particle-body distance.

First, $\bds A$ has to be projected onto the radial, transverse and normal orthogonal unit vectors $\bds{\hat{R}},\bds{\hat{T}},\bds{\hat{N}}$ of the comoving frame of the test particle orbiting the central object.
Their components, in Cartesian coordinates of a reference frame centered in the primary, are  \cite{Monte}
\eqi \bds{\hat{R}} =\left(
       \begin{array}{c}
          \cos\Om\cos u\ -\cos i\sin\Om\sin u\\
          \sin\Om\cos u + \cos i\cos\Om\sin u\\
         \sin i\sin u \\
       \end{array}
     \right)\lb{ierre}
\eqf
 \eqi \bds{\hat{T}} =\left(
       \begin{array}{c}
         -\sin u\cos\Om-\cos i\sin\Om\cos u \\
         -\sin\Om\sin u+\cos i\cos\Om\cos u \\
         \sin i\cos u \\
       \end{array}
     \right)\lb{itrav}
\eqf
\eqi \bds{\hat{N}} =\left(
       \begin{array}{c}
          \sin i\sin\Om \\
         -\sin i\cos\Om \\
         \cos i\\
       \end{array}
     \right)\lb{inorm}.
\eqf
In \rfr{ierre}-\rfr{inorm}, \textcolor{black}{the angles $\Om,u,i$ are as follows. The angle $\Om$ is the longitude of the ascending node: it lies in the reference $\{x,y\}$ plane from the reference $x$ direction to  the intersection of the orbital plane with the reference plane $\{x,y\}$ (the line of the nodes). The angle $u\doteq \omega + f$ is the argument of latitude. In it, $\omega$ is the argument of pericenter: it is an angle in the orbital plane reckoned from the line of the nodes to the point of closest approach, generally known as pericenter. The angle $f$ is the true anomaly: lying in the orbital plane, it is counted from the pericenter to the instantaneous position of the test particle. The angle $i$ is the inclination of the orbital plane to the reference $\{x,y\}$ plane}. In this specific case, we will choose the unit vector $\bds{\hat{\rho}}$ of the line-of-sight, pointing from the object to the observer, to be  directed along the positive $z$ axis, so that the $\{x,y\}$ plane coincides with the usual plane of the sky which is tangential to the celestial sphere at the position of the BH. With such a choice, corresponding to the frame actually used in data reduction \cite{Eise,Ghez}, $i$ is  the inclination of the orbital plane to the plane of the sky ($i=90$ deg corresponds to edge-on orbits, while $i=0$ deg implies face-on orbits), and $\Om$ is an angle in it counted from  the reference $x$ direction; it is such a node which is actually determined from the observations \cite{Ghez,Gille,Gille2}, and, in general, it is not referred to the SBH's equator.
Subsequently, the projected components of $\bds A$ have to be evaluated onto the Keplerian ellipse
\eqi r=\rp{p}{1+e\cos f},\ p\doteq a(1-e^2),\lb{rkep}\eqf where $p$ is the semilatus rectum and $a,e$ are the semimajor axis and the eccentricity, respectively.
The Cartesian coordinates of the Keplerian motion in space are
 \cite{Monte}
 \begin{equation}
{\begin{array}{lll}
 x &=& r\left(\cos\Om\cos u\ -\cos i\sin\Om\sin u\right),\\  \\
 y &=& r\left(\sin\Om\cos u + \cos i\cos\Om\sin u\right),\\  \\
 z &=& r\sin i\sin u.
\end{array}}\lb{xyz}
 \end{equation}
 %The cartesian components of the velocity can be obtained as
 %\eqi
 %\begin{array}{lll}
 %v_x &=& \derp{x}{f}\dert{f}{t},\\ \\
 %
 %v_y &=& \derp{y}{f}\dert{f}{t},\\ \\
 %
 %v_z &=& \derp{z}{f}\dert{f}{t},\\ \\
 %\end{array}
 %\eqf
%in which $df/dt$ is given by
%\cite{Roy}
% \eqi dt = \rp{(1-e^2)^{3/2}}{n(1+e\cos f)^2}df,\lb{yga}\eqf
% where  $n\doteq \sqrt{GM/a^3}$ is the Keplerian mean motion  related to the orbital period by $n=2\pi/P_{\rm b}$.
% Thus, they are
%\eqi
%{ \begin{array}{lll}
% v_x &=& -\rp{an\left[\cO\left(\su+e\so\right)+\ci\sO\left(\cu+e\co\right)\right]}{\sqrt{1-e^2}},\\ \\
 %
% v_y &=& \rp{an\left[-\sO\left(\su+e\so\right)+\ci\cO\left(\cu+e\co\right)\right]}{\sqrt{1-e^2}},\\ \\
 %
% v_z &=& \rp{an\si\left(\cu+e\co\right)}{\sqrt{1-e^2}}.\\ \\
% \end{array}}\lb{vxvyz}
% \eqf

Then,  $A_R,A_T,A_N$ are to be \textcolor{black}{plugged} into the right-hand-sides of the  Gauss equations for the variations of the osculating Keplerian orbital elements \cite{Roy,Soffel}.
%
%
%\begin{widetext}
%\begin{equation}
%{\begin{array}{lll}
%\dert{a}{t}&=& \rp{2}{n\sqrt{1-e^2}}\left[A_R e \sin f+ A_T\left(\rp{p}{r}\right)\right],\\ \\
%
%\dert{e}{t}&=& \rp{\sqrt{1-e^2}}{na}\left\{ A_R\sin f +A_T\left[\cos f+\rp{1}{e}\left(1-\rp{r}{a}\right)\right]\right\},\\ \\
%
%\dert{i}{t}&=&\rp{1}{na\sqrt{1-e^2}}A_N\left(\rp{r}{a}\right)\cos u,\\\\
%
%\dert{\Om}{t}&=& \rp{1}{na\sqrt{1-e^2}\sin i}A_N\left(\rp{r}{a}\right)\sin u,\\\\
%
%\dert{\omega}{t}&=& -\cos i\dert{\Om}{t}+\rp{\sqrt{1-e^2}}{nae}\left[-A_R\cos f+A_T\left(1+\rp{r}{p}\right)\sin f\right], \\ \\
%
%\dert{\varpi}{t}&=& 2\sin^2\left(\rp{i}{2}\right)\dert{\Om}{t}+\rp{\sqrt{1-e^2}}{nae}\left[-A_R\cos f+A_T\left(1+\rp{r}{p}\right)\sin f\right], \\ \\
%
%\dert{\mathcal{M}}{t}&=& n-\rp{2}{na}A_R\left(\rp{r}{a}\right)-\rp{(1-e^2)}{nae}\left[-A_R\cos f+A_T\left(1+\rp{r}{p}\right)\sin f\right],
%\end{array}}\lb{Gauss}
% \end{equation}
% \end{widetext}
% where \textcolor{black}{$n\doteq \sqrt{GM/a^3}$ is the Keplerian mean motion  related to the orbital period by $n=2\pi/P_{\rm b}$}, $\mathcal{M}$ is the mean %anomaly, and $\varpi\doteq\omega+\Om$ is the longitude of pericenter: it is a \virg{dogleg} angle.

\textcolor{black}{Their right-hand-sides}, computed for the perturbing accelerations of the dynamical effect considered, have to be inserted into the analytic expression of the time variation $d\Ps/dt$ of the osculating Keplerian orbital element $\Ps$  of interest. Then, it must be averaged over one orbital revolution by means of
\textcolor{black}{\citep{Roy}
 \eqi dt = \rp{(1-e^2)^{3/2}}{n(1+e\cos f)^2}df\lb{yga},\eqf
where $n\doteq \sqrt{GM/a^3}$ is the Keplerian mean motion  related to the orbital period by $n=2\pi/P_{\rm b}$, }
\textcolor{black}{to obtain $\left\langle d\Ps/dt\right\rangle$. As a general remark, we point out that it would be incorrect to make inferences about the averaged orbital effects $\left\langle d\Ps/dt\right\rangle$ from a simple inspection of the analytic form of the components $A_R,A_T,A_N$ of a given perturbing acceleration $\bds A$, apart from simple trivial cases. The actual calculation must  ultimately be performed in full as previously outlined. Indeed, it may well happen that nonzero components of $\bds A$ yield vanishing averaged variations $\left\langle d\Ps/dt\right\rangle$ for some Keplerian orbital elements $\Ps$. A trivial case occurs, of course, when one or more components of $\bds A$ are identically zero. Conversely, it would be incorrect to argue that certain components of $\bds A$ should necessarily vanish only because the averaged variations of the orbital elements involving them have been found to be zero. Moreover, simple back-\textcolor{black}{of}-the-envelope numerical evaluations of the size of the averaged variation $\left\langle d\Ps/dt\right\rangle$ of a given Keplerian orbital element $\Ps$ which are based on the order of magnitude of $A$ may be misleading as well. Indeed, it may happen that the final result $\left\langle d\Ps/dt\right\rangle$ of the complete calculation  retains a multiplicative factor $e^\alpha, \alpha=\pm 1,\pm2,\pm3,\ldots$ which can cause a notable quantitative difference with respect to what naively guessed, especially for low-eccentricity systems.}
 \section{Calculation of the long-term orbital effects}\lb{orbisa}
 \subsection{The long-term precessions caused by the quadrupole mass moment of the central body for an arbitrary orientation of its spin axis}\lb{tre}
 The  acceleration experienced by a test particle orbiting a nonspherical central mass rotating about a generic direction $\kap$ is
\eqi\bds A^{(Q_2)} = \rp{3Q_2 }{2r^4}\left\{\left[1-5\left(\bds{\hat{r}}\bds\cdot\kap\right)^2\right]\bds{\hat{r}}+2\left(\bds{\hat{r}}\bds\cdot\kap\right)\kap\right\},\lb{accelQ}\eqf
where $Q_2$ is the quadrupole moment of the body, with $[Q_2]={\rm L}^5 {\rm T}^{-2}$. A dimensionless quadrupole parameter $J_2$ can be introduced by posing $Q_2\rightarrow -GM  \mathcal{R}_e^2 J_2$, where $\mathcal{R}_e$ is the equatorial radius of the rotating body.
According to the \virg{no-hair} or uniqueness theorems of GTR \cite{hair1,hair2}, an electrically neutral BH is completely characterized by its mass $M$ and angular
momentum $S$ only. As a consequence, all the multipole moments
of its external spacetime are functions of $M$ and $S$ \cite{multi1,multi2}. In particular,
the quadrupole  moment of the BH is
\eqi Q_2=-\rp{S^2 G}{c^2 M}.\lb{hair}\eqf
The spatial orientation of the BH's spin axis can be considered as unknown. Thus, looking for a more direct connection with actually measurable quantities, we will retain a generic orientation for $\kap$ in the ongoing calculation, i.e., we will not align it to any of  axes of the reference frame used.
After having computed the $R-T-N$ components of \rfr{accelQ} by means of \rfr{ierre}-\rfr{inorm} as
\begin{equation}
{\begin{array}{lll}
A_R^{(Q_2)}=\bds{A}^{(Q_2)}\cdot\bds{\hat{R}}, \\ \\
A_T^{(Q_2)}=\bds{A}^{(Q_2)}\cdot\bds{\hat{T}}, \\ \\
A_N^{(Q_2)}=\bds{A}^{(Q_2)}\cdot\bds{\hat{N}},
\end{array}}\lb{GaussQ}
 \end{equation}
 to be evaluated onto the unperturbed Keplerian ellipse,
it is possible to obtain
\eqi\textcolor{black}{\left\langle\dert a t\right\rangle =\left\langle\dert e t\right\rangle =0,\lb{prima}}\eqf for the semimajor axis and the eccentricity, as in the standard calculations \textcolor{black}{\cite{Roy}} in which $\kap$ is usually aligned with the $z$ axis.

Instead, the inclination $i$ undergoes a long-term variation given by
\eqi\textcolor{black}{\left\langle \dert i t\right\rangle} = \rp{3Q_2}{2\sqrt{GMa^7}\left(1-e^2\right)^2}\mathfrak{I}\left(\Om,i;\kap\right),\lb{Qstrazio}\eqf
with
\begin{widetext}
\eqi\mathfrak{I}\left(\Om,i;\kap\right)\doteq\left(\kx\cO+\ky\sO\right)\left[\kz\ci+\si\left(\kx\sO-\ky\cO\right)\right].\eqf
\end{widetext}
If $\kx=\ky=0$, as in the usual calculation \cite{Roy}, $i$ stays constant.

Concerning the node $\Om$, its long-term variation is
\eqi\textcolor{black}{\left\langle \dert \Om t\right\rangle} = -\rp{3Q_2}{4\sqrt{GMa^7}\left(1-e^2\right)^2}\mathfrak{O}\left(\Om,i;\kap\right),\eqf
with
\begin{widetext}
\eqi
%\begin{array}{lll}
\mathfrak{O}\left(\Om,i;\kap\right)\doteq  2\kz\cos 2i\csc i\left(\kx\sO-\ky\cO\right)
+\ci\left[\kx^2+\ky^2-2\kz^2+\left(\ky^2-\kx^2\right)\cos 2\Om - 2\kx\ky\sin 2\Om\right]\lb{omegone}.
%\end{array}
\eqf
\end{widetext}
Notice that $\kx=\ky=0$ in \rfr{omegone} yields the standard secular precession \cite{Roy} with \eqi\mathfrak{O}\left(i\right)=-2\ci.\eqf

The long-term change of the argument of pericenter $\omega$ is a little more cumbersome. It is
\eqi \textcolor{black}{\left\langle\dert\omega t\right\rangle} = -\rp{3Q_2}{16\sqrt{GMa^7}\left(1-e^2\right)^2}\mathfrak{o}\left(\Om,i;\kap\right),\eqf
with
\begin{widetext}
\eqi
\begin{array}{lll}
\mathfrak{o}\left(\Om,i;\kap\right)&\doteq & 8-11\kx^2-11\ky^2-2\kz^2+\left(\ky^2-\kx^2\right)\cos 2\Om-\\ \\
&-& 2\kz\left(\cot i -5\cos 3 i\csc i\right)\left(\ky\cO-\kx\sO\right)-2\kx\ky\sin 2\Om+\\\\
&+&5\cos 2 i\left[2\kz^2-\kx^2-\ky^2+\left(\kx^2-\ky^2\right)\cos 2\Om+2\kx\ky\sin 2\Om\right].\lb{strazio}
\end{array}
\eqf
\end{widetext}
In the case $\kx=\ky=0$ \rfr{strazio} reduces to \eqi\mathfrak{o}\left(i\right)=2\left(3+5\cos 2 i\right)=4\left(4-5\sin^2 i\right),\eqf which yields the standard expression for the secular precession of the pericenter \cite{Roy}.

The longitude of the pericentre \textcolor{black}{$\varpi\doteq\omega+\Om$, which is a \virg{dogleg} angle,} experiences a long-term variation given by
\eqi \textcolor{black}{\left\langle\dert\varpi t\right\rangle} = -\rp{3Q_2}{16\sqrt{GMa^7}\left(1-e^2\right)^2}\mathfrak{V}\left(\Om,i;\kap\right),\eqf
with
\begin{widetext}
\eqi
\begin{array}{lll}
\mathfrak{V}\left(\Om,i;\kap\right)&\doteq &

8-11\kx^2-11\ky^2-2\kz^2 +\left(\kx^2+\ky^2-2\kz^2\right)\left(4\ci-5\cos 2 i\right)-\\ \\
&-& 4\left(\kx^2-\ky^2\right)\left(3+5\ci\right)\sin^2\left(\rp{i}{2}\right)\cos 2\Om- 2\ky\kz\sec\left(\rp{i}{2}\right)\left[\sin\left(\rp{3i}{2}\right) +5\sin\left(\rp{5i}{2}\right)\right]\cO +\\ \\
&+& 2\kx\kz\sec\left(\rp{i}{2}\right)\left[\sin\left(\rp{3i}{2}\right) +5\sin\left(\rp{5i}{2}\right)\right]\sO - 8\kx\ky\sin^2\left(\rp{i}{2}\right)\left(3+5\ci\right)\sin 2\Om.
\lb{macello}
\end{array}
\eqf
\end{widetext}
For $\kx=\ky=0$ \rfr{macello} reduces to
\eqi \mathfrak{V}\left(i\right)= 2\left[3-\left(4\ci -5\cos 2 i\right)\right]=4(4-5\sin^2 i-2\ci),\eqf which yields
the usual expression for the secular rate of $\varpi$ \cite{Roy}.

Finally, the long-term change of the mean anomaly at epoch \textcolor{black}{$\mathcal{M}_0\doteq n(t_0 - t_{\rm p})$, where $t_{\rm p}$ is time of passage at pericenter,}  is
\eqi\textcolor{black}{\left\langle\dert{\mathcal{M}_0} t\right\rangle} =\rp{3Q_2}{16\sqrt{GMa^7\left(1-e^2\right)^3}}\mathfrak{M}\left(\Om,i;\kap\right), \eqf
with
\begin{widetext}
\eqi
\begin{array}{lll}
\mathfrak{M}\left(\Om,i;\kap\right)&\doteq &
-8+9\kx^2+9\ky^2+6\kz^2+3\left(\kx^2+\ky^2-2\kz^2\right)\cos 2 i +6\left(\kx^2-\ky^2\right)\sin^2 i\cos 2\Om+\\ \\
&+& 12\left[\kz\sin 2 i\left(\ky\cO-\kx\sO\right)+\kx\ky\sin^2 i\sin 2\Om\right].\lb{Mstrazio}
\end{array}
\eqf
\end{widetext}
Also in this case, for $\kx=\ky=0$ the standard secular precession \cite{Roy} is recovered since \rfr{Mstrazio} reduces to
\eqi \mathfrak{M}\left(i\right)=-2\left(1+3\cos 2 i\right)=-4\left(2-3\sin^2 i\right).\eqf

Incidentally, we remark that the field of applicability of  \rfr{prima}-\rfr{Mstrazio} is not limited just to the BH arena, being \textcolor{black}{then} generally valid also for astrophysical binary systems, stellar planetary systems, and planetary satellite geodesy. In particular, they could be useful when satellite-based tests of GTR are performed or designed (See Sec. V).
\subsection{The Lense-Thirring long-term precessions for a generic orientation of the spin axis of the central body}\lb{quattro}
According to GTR, the gravitomagnetic Lense-Thirring acceleration felt by a test particle moving with velocity $\bds v$ around a rotating body with angular momentum $\bds S=S\kap$ at large distance from it is
\eqi\bds A^{(\rm LT)}=-2\left(\rp{\bds v}{c}\right)\bds\times\bds B_g.\lb{accelLT}\eqf
In \rfr{accelLT} the gravitomagnetic field $\bds B_g$, far from the central object where the Kerr metric \cite{Kerr}  reduces to the Lense-Thirring one, is
\eqi\bds B_g=-\rp{GS}{cr^3}\left[\kap-3\left(\kap\bds\cdot\bds{\hat{r}}\right)\bds{\hat{r}}\right].\lb{Bg}\eqf
Concerning $S$, the existence of the horizon in the Kerr metric \cite{Kerr}  implies a maximum value for the angular momentum of a spinning BH \cite{Bar1,Mel1}, so that
$ S=\chi_g S_{\rm max},$
with\eqi S_{\rm max} = \rp{M^2 G}{c}.\lb{limit}\eqf
If $\chi_g>1$,  the Kerr metric
\cite{Kerr}  would have a naked singularity without a
horizon. Thus, closed timelike curves could be considered, implying a causality violation \cite{Chan}.
Although not yet  proven, the cosmic censorship
conjecture \cite{Pen69} asserts that naked
singularities cannot be formed via the gravitational collapse
of a body.
If the limit of \rfr{limit} is actually reached or not by  astrophysical BHs depends on their accretion history \cite{Barde}. In the case of Sgr A$^{\ast}$, it may be $\chi_g\approx 0.44-0.52$ \cite{Genznat,Kato} or even less \cite{spin2,mmvlbi2}. Contrary to BHs,
 no theoretical constraints on the value of $\chi_g$  exist for stars.
For main-sequence
stars, $\chi_g$ depends sensitively on the stellar mass, and can be
much larger than unity \cite{Kraft1,Kraft2,Dicke,stella}. The case of compact stars was recently treated in Ref. \cite{cinesi}, showing that for neutron stars with $M\gtrsim 1 M_{\odot}$ it should be $\chi_g\lesssim 0.7$, independently of the Equation Of State (EOS) governing the stellar matter. Hypothetical quark stars may have $\chi_g>1$, strongly depending on the EOS and the stellar mass \cite{cinesi}.

In  the standard derivations of the Lense-Thirring effect \cite{LT} existing in literature the reference $\{x,y\}$ plane was usually chosen coincident with the equatorial plane of the rotating mass.
In principle, the Lense-Thirring orbital precessions for a generic orientation of $\bds S$ could be worked out with the Gauss equations in the same way as done for $Q_2$. Anyway, they were recently worked out \cite{Iorio010}, in a different framework, with the less cumbersome Lagrange planetary equations \cite{Roy}.
For the \textcolor{black}{reader's convenience}, we display here the final result
\begin{widetext}
\begin{equation}
\begin{array}{lll}
\textcolor{black}{\left\langle \dert a t\right\rangle}  & = & 0, \\ \\
\textcolor{black}{\left\langle\dert e t\right\rangle}  & = & 0, \\ \\
\textcolor{black}{\left\langle\dert i t\right\rangle}  & = & \rp{2GS\left(\kx \cos\Om+\ky\sin\Om\right)}{c^2 a^3(1-e^2)^{3/2}}, \\ \\
\textcolor{black}{\left\langle\dert\Om t\right\rangle}  & = & \rp{2GS\left[ \kz + \cot i\left(\ky\cos\Om -\kx \sin\Om\right)\right]}{c^2 a^3(1-e^2)^{3/2}}, \\ \\
\textcolor{black}{\left\langle\dert\omega t\right\rangle}  & = & -\rp{GS\left[6\kz\ci +\left(3\cos 2 i -1\right)\csc i\left(\ky\cO-\kx\sO\right)\right]}{c^2 a^3(1-e^2)^{3/2}}, \\ \\
\textcolor{black}{\left\langle\dert\varpi t\right\rangle}  & = & -\rp{ GS \left\{4\left[\kz\cos i+\sin i\left(\kx\sin\Om-\ky\cos\Om \right)  \right]
-2\left[\kz\sin i+\cos i\left(\ky\cos\Om-\kx\sin\Om \right) \right]\tan(i/2)
\right\}}{c^2 a^3(1-e^2)^{3/2}}, \\ \\
\textcolor{black}{\left\langle\dert{\mathcal{M}} t\right\rangle} & = & 0.
\end{array}\lb{piccololt}
\end{equation}
\end{widetext}
 Notice that \rfr{piccololt}  yields just the usual Lense-Thirring secular rates \cite{LT, Soffel} for $\kx=\ky=0$.
  %concerning $\varpi$, by posing $i/2\doteq \zeta$ it can be shown that $-2\cos i + \sin i\tan (i/2)=1-3\cos i$.
  Contrary to such a scenario, the inclination $i$ experiences a long-term gravitomagnetic change for an arbitrary orientation of $\bds S$: it is independent of the inclination $i$ itself.
 %Moreover, $S_x$ and $S_y$ induce additional precessions for the node $\Om$, the argument of pericentre $\omega$ and the longitude of pericentre $\varpi$ which %depend on $i$. All the additional precessions depend on $\Om$ as well.
%
\subsection{A comparison with a different approach}\lb{compa}
Will \cite{Will08} refers $\kap$ to the orbital plane of a generic star by choosing $\bds{e}_p,\bds{e}_q,\bds{h}$ as orthonormal vectors: $\bds{e}_p$ is directed along the line of the nodes, $\bds{e}_q$ lies in the orbital plane perpendicularly to $\bds{e}_p$, and $\bds h$ is directed along the orbital angular momentum.
%They are
%
%
%\eqi \bds{e}_p =\left(
%       \begin{array}{c}
%          \cos\Om\\
%          \sin\Om\\
%         0 \\
%       \end{array}
%     \right)\lb{ip}
%\eqf
%
%
% \eqi \bds{e}_q =\left(
%       \begin{array}{c}
%         -\cos i\sin\Om \\
%         \cos i\cos\Om \\
%         \sin i \\
%       \end{array}
%     \right)\lb{iq}
%\eqf
%\eqi \bds{h} =\left(
%       \begin{array}{c}
%          \sin i\sin\Om \\
%         -\sin i\cos\Om \\
%         \cos i\\
%       \end{array}
%     \right)\lb{ih}.
%\eqf
%
%
%
\textcolor{black}{The unit vectors $\bds{e}_p$ and $\bds{e}_q$  can be obtained from  \rfr{ierre} and \rfr{itrav}, respectively, by posing $u\rightarrow 0$, while $\bds h$ coincides with \rfr{inorm}.} Thus, one has
\begin{equation}
\begin{array}{lll}
\kx &=& \kp\cO+\left(\kh\si-\kq\ci\right)\sO, \\ \\
\ky &=& \kp\sO-\left(\kh\si-\kq\ci\right)\cO, \\ \\
\kz &=& \kh\ci+\kq\si, \\ \\
\end{array}\lb{kpkqkh}
\end{equation}
\textcolor{black}{where $\kp,\kq,\kh$ can straightforwardly be expressed in terms of the polar angles $\alpha$ and $\beta$ used by Will \cite{Will08} in the orbital frame.}
%
%\begin{equation}
%\begin{array}{lll}
%\kp &=& \sa\cb, \\ \\
%
%\kq &=& \sa\sb, \\ \\
%
%\kh &=& \ca,
%
%\end{array}\lb{kkk}
%\end{equation}
%having introduced

Inserting \rfr{kpkqkh}
%and \rfr{kkk}
into the equations of Sec. \ref{tre} and Sec. \ref{quattro} allows us to obtain Eq. (2a), Eq. (2b), and Eq. (2c) of Ref. \cite{Will08} after some  algebra.
\subsection{Stellar orbital perturbations caused by ultralow frequency gravitational waves}\lb{onda}
The stars orbiting the SBH in Sgr A$^{\ast}$ could also be  used, in principle, as probes for detecting or constraining plane gravitational waves of ultralow frequency ($\nu\approx 10^{-8}-10^{-10}$ Hz or less) impinging on the system from the outside. Indeed, the passage of such  waves through the orbits of the closest stars
would cause long-term variations of all their Keplerian orbital elements, apart from the semimajor axis $a$. They have recently been worked out in Ref. \cite{Iorio011b} for general orbital configurations, i.e., without making a-priori assumptions on their inclinations and eccentricities of the perturbed test particle, and arbitrary  directions of incidence for the wave.
Conversely, gravitational waves can be generated within the stellar system of Sgr A$^{\ast}$, as discussed in Ref. \cite{Freitag}.
\section{Numerical evaluations}\lb{cinque}
In Table \ref{tavola} we quote the relevant physical and orbital parameters for the SBH-S2 system. The orbital period of S2 is $P_{\rm b}=15.98$ yr, so that the astrometric measurements currently available cover a full revolution of it.
\begin{table*}
\caption{\label{tavola}Relevant physical and orbital parameters of the SBH-S2 system in Sgr A$^{\ast}$ (first row), and their uncertainties (second row). The Keplerian orbital elements of S2 were retrieved from Table 1 of  \protect{\cite{Gille}}. The figure for $\chi_g$ comes from Ref. \protect{\cite{Genznat}}, while the one for the gravitational parameter $\mu\doteq GM$ is from a multi-star fit yielding $\mu=(4.30\pm 0.50)\times 10^6\mu_{\odot}$ \protect{\cite{Gille}}. The quoted value in m for the semimajor axis of S2  was obtained by multiplying  its angular value  $a=0.1246\pm 0.0019$ arcsec \protect{\cite{Gille}} by the distance of the SBH $d=8.28\pm 0.44$ kpc \protect{\cite{Gille}}. For the angular momentum and the quadrupole moment of the SBH we used $S=\chi_g (M^2 G)/c $ and $Q_2=-(S^2 G)/(c^2 M)=-\chi^2_g(G^3M^3)/c^4$. The orbital period of S2 is $P_{\rm b}=15.98\ {\rm yr}=5.04\times 10^8$ s. The figures for $S$ and $Q_2$, obtained in the hypothesis that GTR is correct, strongly depend on $\chi_g$, which is, at present, highly uncertain. For example, the authors of Ref. \protect{\cite{Kato}} yield $\chi_g=0.44\pm 0.08$, while for the authors of Ref. \protect{\cite{spin2,mmvlbi2}} it could be even smaller. We will use them  to indicatively give order-of-magnitude evaluations of the additional orbital precessions which would occur because of $S$ and $Q_2$ according to GTR.
}
\begin{ruledtabular}
\begin{tabular}{llllllll}
$\mu$ (m$^3$ s$^{-2}$) & $S$ (kg m$^2$ s$^{-1}$) & $Q_2$ (m$^5$ s$^{-2}$) &  $\chi_g$ & $a$ (m) & $e$  & $i$ (deg) & $\Om$ (deg) \\
\colrule
$5.70\times 10^{26}$ & $8.46\times 10^{54}$ & $-6.22\times 10^{45}$ &  $0.52$ & $1.54\times 10^{14}$ & $0.8831$ & $134.87$ & $226.53$\\
$6.6\times 10^{25}$ & $4.66\times 10^{54}$ & $6.58\times 10^{45}$ & $ 0.26$  & $8\times 10^{12}$ & $0.0034$ & $0.78$ & $0.72$ \\
\end{tabular}
\end{ruledtabular}
\end{table*}

The quadrupole-induced precessions of \rfr{Qstrazio}-\rfr{Mstrazio} are all linear combinations of the products of the components of $\kap$ plus, sometimes,  a term independent of $\kap$: they can be cast into the form
\begin{widetext}
\eqi \textcolor{black}{\left\langle\dert\Ps t\right\rangle} = D_0\left(Q_2,a,e,i,\Omega\right)+\rp{1}{2}\sum_{s,l} D_{sl}\left(Q_2,a,e,i,\Omega\right) \hat{k}_s\hat{k}_l,\ s,l=x,y,z,\ \Ps=i,\Omega,\omega,{\mathcal{M}}.\eqf
\end{widetext}
The numerical values of the coefficients $D_0$ and $D_{sl}=D_{ls}$ for S2, in $\mu$as yr$^{-1}$, are quoted in Table \ref{tavolaQ2}.
\begin{table}
\caption{\label{tavolaQ2}Coefficients of the quadrupole precessions of S2, in $\mu$as yr$^{-1}$, according to Table \protect{\ref{tavola}}. GTR was assumed for $Q_2$, with $\chi_g=0.52$.
}
\begin{ruledtabular}
\begin{tabular}{llllllll}
 & $D_0$ &  $D_{x^2}$ & $D_{y^2}$ & $D_{z^2}$ & $D_{xy}$ & $D_{xz}$ & $D_{yz}$ \\
\colrule
$i$ & $0$ & $406$ & $-406$ & $0$ & $43$ & $558$ & $588$ \\
$\Omega$ & $0$ & $427$ & $384$ & $-810$ &  $-809$ & $-5$ & $5$\\
$\omega$ & $-1149$ & $1568$ & $1584$ & $294$ &  $293$ & $-1254$ & $1189$\\
$\mathcal{M}$ & $-539$ & $595$ & $616$ & $406$ & $405$ & $-587$ & $556$ \\
\end{tabular}
\end{ruledtabular}
\end{table}
The largest effects occur for $\omega$ and $\mathcal{M}$ because of $D_0$, which is of the order of $\approx 1$ milliarcsec yr$^{-1}$ (mas yr$^{-1}$). The other terms are damped by the square of the components of $\kap$. Moreover, partial mutual \textcolor{black}{cancellation} may occur depending on the orientation of the SBH spin axis.

The Lense-Thirring precessions of \rfr{piccololt} are all linear combinations of the components of $\kap$: they can be cast into the form
\eqi \textcolor{black}{\left\langle\dert\Ps t\right\rangle} = \sum_j C_j\left(S,a,e,i,\Omega\right) \hat{k}_j,\ j=x,y,z,\ \Ps=i,\Omega,\omega.\eqf
The numerical values of the coefficients $C_j$ for S2, in arcsec yr$^{-1}$, are listed in Table \ref{tavolaLT}.
\begin{table}
\caption{\label{tavolaLT}Coefficients of the Lense-Thirring precessions of S2, in arcsec yr$^{-1}$, according to Table \protect{\ref{tavola}}. In particular, $\chi_g=0.52$ was used for the spin of the SBH.
}
\begin{ruledtabular}
\begin{tabular}{llll}
& $C_x$ & $C_y$ & $C_z$ \\
\colrule
$i$ & $-0.14$ & $-0.15$ & $0$ \\
$\Omega$ & $-0.15$ & $0.14$ & $0.21$ \\
$\omega$ & $0.11$ & $-0.10$ & $0.45$ \\
\end{tabular}
\end{ruledtabular}
\end{table}
They are of the order of about $10^{-1}$ arcsec yr$^{-1}$, i.e.,  orders of magnitude larger than the quadrupole precessions of Table \ref{tavolaQ2}: also in this case, partial mutual \textcolor{black}{cancellations} may occur depending on $\kap$, thus impacting the detectability of the gravitomagnetic rates.

The figures of Table \ref{tavolaQ2} and Table \ref{tavolaLT} can be compared with the present-day accuracies in empirically determining the orbital precessions of S2 listed in Table \ref{tavola_errori}.
\begin{table}
\caption{\label{tavola_errori}Naive evaluations of the uncertainties in the secular variations of  the S2 osculating Keplerian orbital elements, in arcsec yr$^{-1}$, obtained by dividing the errors in the elements from Table 1 of Ref. \protect{\cite{Gille}} by a time interval $\Delta T\approx P_{\rm b}$.
Concerning the mean anomaly, its uncertainty was evaluated from that of the time of periastron passage $t_{\rm p}$, released in Ref. \protect{\cite{Gille}}, according to the expression for the mean anomaly at the epoch of periastron passage $\mathcal{\mathcal{M}}_0=-n t_{\rm p}$; also the errors coming from $a$ and $\mu$ through $n$ were taken into account.
}
\begin{ruledtabular}
\begin{tabular}{llll}
$\sigma_{\dot{i}}$  & $\sigma_{\dot\Omega}$  & $\sigma_{\dot\omega}$  & $\sigma_{\dot{\mathcal{M}}}$  \\
\colrule
$176$ & $163$ &  $182$ & $1203$ \\
\end{tabular}
\end{ruledtabular}
\end{table}
They are of the order of $10^2-10^3$ arcsec yr$^{-1}$. The Lense-Thirring precessions of S2 (Table \ref{tavolaLT}) are about three orders of magnitude smaller than the current accuracy, while the quadrupole effects of Table \ref{tavolaQ2} are negligibly small.

By considering a fictitious  star $X$  with, say, the same orbital parameters of S2 apart from the semimajor axis $a$, assumed to be one order of magnitude smaller so that its orbital period would  just be $P_{\rm b}=0.5$ yr, it turns out that its 1PN GTR periastron precession would be as large as 4 deg yr$^{-1}$, while its Lense-Thirring and quadrupole precessions would be of the order of about $\approx 10^2$ arcsec yr$^{-1}$ and $\approx 1$ arcsec yr$^{-1}$, respectively.

If, as expected, angular shifts of $\Delta\xi\approx 10$ $\mu$as, as seen from the Earth,  will  really become measurable in future thanks to GRAVITY and ASTRA,
this would imply an accuracy of the order of $\Delta\Ps\approx \left(d/a\right)\Delta\xi=16$ arcsec for S2, and 160 arsec for a  star one order of magnitude closer to the SBH.  If such targets will be discovered, their Lense-Thirring  shifts should become  detectable after some years, while the $Q_2-$induced perturbations would still remain hard to measure, \textcolor{black}{even} for $e\approx 0.9$.
\section{Summary and conclusions}\lb{sei}
We analytically worked out the long-term, i.e., averaged over one full revolution, variations of all the six osculating Keplerian orbital elements of a test particle orbiting a nonspherical, spinning body endowed with angular momentum $\bds S$ and quadrupole moment $Q_2$ for a generic spatial orientation of its spin axis $\kap$. We  did not restrict ourselves to any specific orbital configuration \textcolor{black}{of the particle}. Here we applied our results to the  stars  orbiting the SBH in Sgr A$^{\ast}$: those identified so far  are  moving  along highly elliptical trajectories with periods $P_{\rm b}\geq 16$ yr.
The current level of accuracy in empirically determining the precessions of the angular orbital elements of S2,  having $P_{\rm b}=16$ yr, can be evaluated to be of the order of $\approx 10^2-10^3$ arcsec yr$^{-1}$. The predicted 1PN GTR periastron precession of S2, which is independent of the orientation of the spin axis of the SBH,  is $40\pm 10$ arcsec yr$^{-1}$. The predicted GTR spin and quadrupole-induced precessions of S2 are of the order of $\approx 10^{-1}$ arcsec yr$^{-1}$ and $\approx 10^2-10^3\mu$as yr$^{-1}$, respectively: they depend on $\kap$, and  partial \textcolor{black}{cancellations} among their components may occur, thus reducing their magnitude.
Concerning hypothetical stars with orbital periods of less than 1 yr, not yet discovered, the 1PN GTR periastron precessions would be as large as some deg yr$^{-1}$, while the $S$ and $Q_2$ effects would be of the order of $\approx 10^2$ arcsec yr$^{-1}$ and $\approx 1$ arcsec yr$^{-1}$, respectively.
Planned improvements of the infrared telescopes used so far  aim to reach an accuracy level of $\approx 10$ $\mu$as at best in measuring angular shifts as seen from the Earth corresponding to stellar orbital shifts of about $1.6\times 10^1-10^2$ arcsec for S2 and stars closer than it  by one order of magnitude, respectively.
\textcolor{black}{Finally, we stress that our \textcolor{black}{calculations} are not restricted to any specific coordinate system. Thus, in the form in which we obtained them, they  can fruitfully be used also in other  scenarios like astrophysical binaries, stellar planetary systems and planetary satellite geodesy in which tests of post-Newtonian gravity may be involved.} \textcolor{black}{Examples of that can be found, e.g., in Ref. \cite{EPL}, pertaining the Lense-Thirring tests with the LAGEOS satellites in the terrestrial gravitational field, and in Ref. \cite{sole} dealing with Mercury and the gravitomagnetic field of the Sun.}
\section*{Acknowledgements}
I thank S. Gillessen for useful correspondence. I am grateful to A. Hees for having pointed out to me an error sign in Eq. (8) and in the correspondence $Q_2\rightarrow J_2$.
 %-----------------------------------------

%----------

\end{document}